\documentclass[preprint]{revtex4-1}

\usepackage{amsmath}
\usepackage{graphicx}
\usepackage{natmove}

\renewcommand{\v}{\vec}

\begin{document}

\title{Contact area of rough spheres: Large scale simulations and simple scaling laws}

\author{Lars Pastewka}
\email{lars.pastewka@kit.edu}
\affiliation{Institute for Applied Materials, Karlsruhe Institute of Technology, Engelbert-Arnold-Stra\ss e 4, 76131 Karlsruhe, Germany}
\affiliation{Department of Physics and Astronomy, Johns Hopkins University, 3400 North Charles Street, Baltimore, MD 21218, USA}
\author{Mark O. Robbins}
\email{mr@pha.jhu.edu}
\affiliation{Department of Physics and Astronomy, Johns Hopkins University, 3400 North Charles Street, Baltimore, MD 21218, USA}

\date{\today}


\begin{abstract}
We use molecular simulations to study the nonadhesive and adhesive atomic-scale contact of rough spheres with radii ranging from nanometers to micrometers over more than ten orders of magnitude in applied normal load.
At the lowest loads, the interfacial mechanics is governed by the contact mechanics of the first asperity that touches.
The dependence of contact area on normal force becomes linear at intermediate loads and crosses over to Hertzian at the largest loads.
By combining theories for the limiting cases of nominally flat rough surfaces and smooth spheres, we provide parameter-free analytical expressions for contact area over the whole range of loads.
Our results establish a range of validity for common approximations that neglect curvature or roughness in modeling objects on scales from atomic force microscope tips to ball bearings.
\end{abstract}

\maketitle

Contact, friction and adhesion control the behavior of mechanical devices from cars~\cite{Holmberg2012} to artificial joints~\cite{Dowson2012} and determine the properties of granular matter such as colloids or sand piles~\cite{Jaeger:1996p4569}.
The canonical geometry used to measure these quantities is a sphere of radius $R$ pushed into a flat substrate with a normal force $N$.
Recent metrology devices with this geometry have the ability to measure relative motion of surfaces with nanometer resolution, including the surface force apparatus ($R\sim10\,\text{mm}$)~\cite{Israelachvili:2010p036601}, colloidal probes ($R \sim 10\,\mu$m)~\cite{Kappl:2002p129},
and atomic force microscopes ($R\sim 10\,\text{nm}$)~\cite{Giessibl:2003p5804}.
The surfaces in these devices generally have small scale roughness superimposed on the macroscopic curvature that can dramatically alter the contact area, friction, tangential stiffness, adhesion and transport properties.
There has been significant recent progress in treating roughness on nominally flat surfaces but not on the interaction between roughness and macroscopic curvature.

In this paper we use molecular simulations to study contact of rough spheres with $R$ from $30\,$nm to $30\,\mu$m.
By combining recent results for rough flat surfaces~\cite{perssonreview,Hyun:2004p026117,Campana:2007p38005,Akarapu:2011p204301,Putignano:2012p973,Prodanov:2013p433,Carbone:2008p2555,Pastewka:2014p3298} with classical results for smooth spheres~\cite{Hertz:1881p156,Johnson:1971p301,Derjaguin:1975p314,Johnson:Book1985,Maugis:Book1999,Maugis:1992p243}, we develop parameter free equations that describe the full variation of the area of molecular contact $A$ with $N$.
$A$ is difficult to obtain from experiments but it plays a central role in determining contact properties.
The variation of $A$ with $N$ determines whether there is adhesion, and the magnitude and geometry of contact area are widely used to calculate stiffness, transport and the friction force $F$.
A simple approximation that appears to be valid in many cases, even down to sliding AFM tips~\cite{Carpick:1996,Schwarz:1997p6987,Lessel2013}, is that of a constant interfacial shear stress $\tau$~\cite{Bowden:Book1950}, giving a friction force $F=\tau A$.

More than a century ago Hertz showed~\cite{Hertz:1881p156}
that contact of a rigid smooth sphere with radius $R$ on an elastic flat of contact modulus $E^*$
occurs in a circle with radius
\begin{equation}
  a_\text{Hertz}=\left(3NR/4E^*\right)^{1/3}.
  \label{aHertz}
\end{equation}
The work of Hertz has been subsequently extended to many other situations, such as shear loading, interfacial friction and adhesion~\cite{Johnson:1971p301,Derjaguin:1975p314,Johnson:Book1985,Maugis:Book1999,Maugis:1992p243}.

Roughness reduces the area in intimate repulsive contact since only the highest asperities are able to touch.
We therefore expect that the true contact area $A$ is less than $A_\text{Hertz}\equiv \pi a_\text{Hertz}^2$ in most situations.
The prevalent theory of contact of a rough sphere on a flat was presented by Greenwood \& Tripp~\cite{Greenwood:1967p153}.
Their analysis is based on the traditional Greenwood-Williamson (GW) model~\cite{Greenwood:1966p300} for nominally flat surfaces that approximates the rough surface by a collection of noninteracting Hertzian contacts of identical radius.
In the last decade, analytical~\cite{perssonreview} and numerical~\cite{Hyun:2004p026117,Campana:2007p38005,Akarapu:2011p204301,Putignano:2012p973} works have revealed important limitations of GW results for total contact area~\cite{Carbone:2008p2555}
and contact geometry~\cite{Hyun:2004p026117,Persson:2008p312001,Campana:2008p354013} motivating us to revisit the contact of rough spheres.

Recent studies of nonadhesive contact between nominally flat surfaces of area $A_0$ all find a linear relation between $A$ and $N$ up to fractional contact areas $A/A_0 \sim 10\%$~\cite{perssonreview,Hyun:2004p026117,Campana:2007p38005,Carbone:2008p2555,Akarapu:2011p204301,Putignano:2012p973,Prodanov:2013p433}.
This implies a constant normal pressure $p_\text{rough}=N/A$ within the contacting regions.
Roughness enters only through the dimensionless root mean square (rms) slope of the surface $h_{\rm rms}^\prime\equiv\langle |\nabla h|^2\rangle^{1/2}$, where $h(\v{r})$ is the height as a function of position $\v{r}$ in the plane.
Numerical studies find 
$p_\text{rough}\equiv h^\prime_\text{rms} E^*/\kappa$,
where the dimensionless constant
$1/\kappa\approx 1/2$ in the continuum, hard-wall limit~\cite{Hyun:2004p026117,Campana:2007p38005,Akarapu:2011p204301,Putignano:2012p973,Prodanov:2013p433}.
The constant $1/\kappa$ becomes smaller for finite range soft repulsion and adhesive interactions.
Negative values are found for strongly adhesive interactions and correspond to macroscopic adhesion or stickiness.\cite{Pastewka:2014p3298}
At high loads, the surfaces are pushed into compliance and $A=A_0$.
Over the whole range of loads, the fractional area of contact $f=A/A_0$  approximately follows an error function~\cite{perssonreview,Manners:2006p600,Prodanov:2013p433,Yastrebov:2015p83},
\begin{equation}
  f(p)
  =
  \text{erf}\left(\sqrt{\pi}p/2p_\text{rough}\right),
  \label{fracarea}
\end{equation}
where $p=N/A_0$ is the nominal pressure pushing the two surfaces together.

Here we extend these calculations to contact of a rough rigid sphere of radius $R$ and a semi-infinite elastic substrate of contact modulus $E^*$.
The results can be mapped to contact of two elastic rough spheres using
a standard mapping~\cite{Johnson:Book1985}.
The rigid surface has height $h_R + h$ where the first term is the average curvature and the second is the random roughness.
As in flat surface studies~\cite{Hyun:2004p026117,Campana:2007p38005,Akarapu:2011p204301,Putignano:2012p973,Prodanov:2013p433}, $h$ has a self-affine fractal form
from lower wavelength $\lambda_s$ to upper wavelength $\lambda_L$
with Hurst exponent $H$~\cite{supplemental}.

The calculations use
an efficient Green's function method~\cite{Stanley:1997p481,Polonsky:1999p206,Campana:2006p075420,Pastewka:2012p075459}
supplemented by
a padding region that prevents any effect from repeating images~\cite{Hockney:1970p86}.
We use the Greens function for an isotropic solid with Poisson ratio $1/2$~\cite{Johnson:Book1985}.
Atoms on the flat elastic substrate are arranged on a square grid with spacing $a_0$.
This can be regarded as the $(100)$ surface of an fcc lattice or a convenient discretization of the continuum equations of elasticity.
We show results for different interfacial interactions with
$\lambda_s=4 a_0$, $\lambda_L=512 a_0$ and $H=0.8$, but simulations with other parameters were carried out with no fundamental differences in the results.
Contact area is obtained from atoms that feel a repulsive load~\cite{Pastewka:2014p3298}.
Contact radii up to 1024 atoms were studied, corresponding to $\sim 0.3\,\mu$m.

\begin{figure*}
  \begin{center}
    \includegraphics[width=16cm]{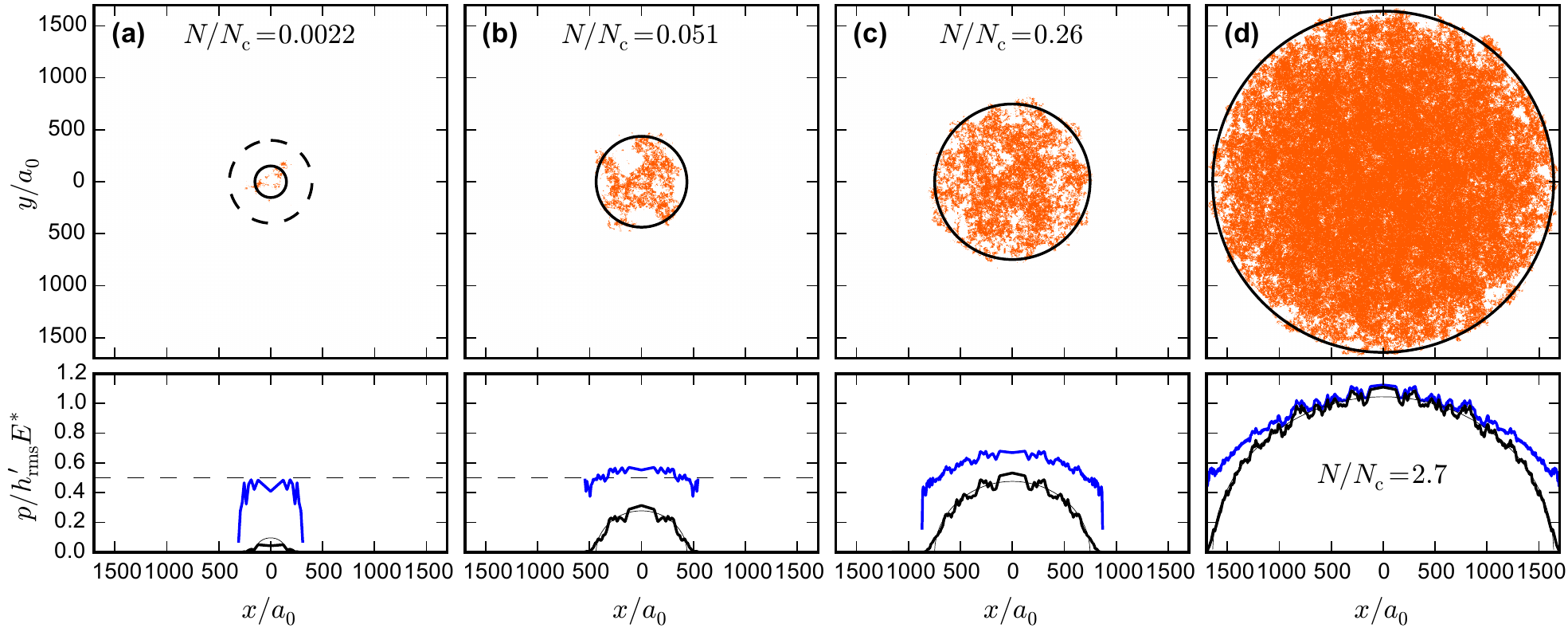}
    \caption{\label{geometry}
	    Hard-wall, nonadhesive contact of a rough sphere of radius $R=10,000 a_0$ as load increases from (a) to (d).
      The top row of panels shows the real contact area (orange) in comparison to the area predicted by Hertz theory (area bounded by the solid line).
      Dashed circle in (a) has radius $a'$ and encloses area of first contact.
      The thick black lines in the bottom row of panels show the radially average pressure distribution $\bar{p}(r)$ and the thin lines show the Hertz prediction.
      %
      The blue lines show $\bar{p}_c$, the pressure averaged over only the orange contacting regions in the top panels.
      The dashed horizonal line shows $p_{rep}$ for two rough flats.
      Roughness has Hurst exponent $H=0.8$ and small and large wavelength cutoffs $\lambda_s=4 a_0$ and $\lambda_L=512 a_0$, respectively.
    }
  \end{center}
\end{figure*}

Fig.~\ref{geometry} shows the result of a nonadhesive calculation with hard-wall interactions between the rough sphere and elastic flat~\cite{Polonsky:1999p206}.
Initial contact occurs at the highest protrusion which is not necessarily located at the bottom of the equivalent smooth sphere.
In the particular case shown in Fig.~\ref{geometry}a, contact occurs outside of the Hertz contact radius $a_\text{Hertz}$ corresponding to the given load (solid black line).

One can estimate the radius $a'$ of the first asperity to contact
by setting the nominal separation between a smooth sphere and substrate
equal to the rms height $h_{\rm rms}$ of the roughness:
$a'^2=2R h_{\rm rms}$.
For $a'<\lambda_L$, $h_\text{rms}$ depends on the scale $a'$ over which it is measured~\cite{aprime}.
Solving for $a'$ gives the analytical expression:
\begin{equation}
a'/\lambda_s = \{2R h^\prime_\text{rms}/\lambda_s [(1-H)/H]^{1/2}]\}^{1/(2-H)}
\end{equation}
This radius is shown by the dashed line in Fig.~\ref{geometry}a.

The bottom row of panels in Fig. 1 shows two measures for the radially averaged pressure in the contact.
The black line shows the full average pressure, $\bar{p}(r)$, that includes noncontacting regions with zero pressure in the average.
As the normal force increases [going from panel (a) to panel (d)] $\bar{p}(r)$ increases, but is close to Hertz's expression (thin solid line) in all cases.
The largest deviation from Hertz's solution are near the contact edge because part of the contacting area sits outside $a_\text{Hertz}$.

Also shown in Fig.~\ref{geometry} is the average contact pressure, $\bar{p}_c(r)$, where in contrast to $\bar{p}(r)$ non-contacting regions are excluded from the average.
For low loads $\bar{p}_c$ is always close to $p_{\rm rough}$ and is independent of radius and load (see Fig.~\ref{geometry}a and \ref{geometry}b).
At larger loads $\bar{p}_c$ rises above $p_{\rm rough}$.
The upper panels of Fig.~\ref{geometry} show that this increase in $\bar{p}_c$ occurs when the surfaces are being pushed into nearly complete contact.
In this limit, $\bar{p}_c$ becomes identical to $\bar{p}$ and both follow the Hertzian form.

The numerical results for the variation of total contact area with load are accurately described by simply assuming that the fraction of the Hertzian area that contacts is given by the result for a nominally flat rough surface at the same mean pressure $N/A_\text{Hertz}$:
\begin{equation}
A/A_\text{Hertz}=f(N/A_\text{Hertz}).
\label{roughsphere}
\end{equation}
While the pressure profile predicted by Hertz varies with $r$ (see Fig.~\ref{geometry}), the associated distribution of local pressures is peaked around the mean pressure $N/A_\text{Hertz}$.
Given this assumption, there should be a crossover from randomly rough to Hertz scaling when $N/A_\text{Hertz}=p_{\rm rough}$ or at a load:
\begin{equation}
N_{\rm c}
=
\left(9\pi^3/16\right) E^* R^2
\left(h_{\rm rms}^\prime/\kappa \right)^3
\label{areacross}
\end{equation}

\begin{figure*}
  \begin{center}
    \includegraphics[width=16cm]{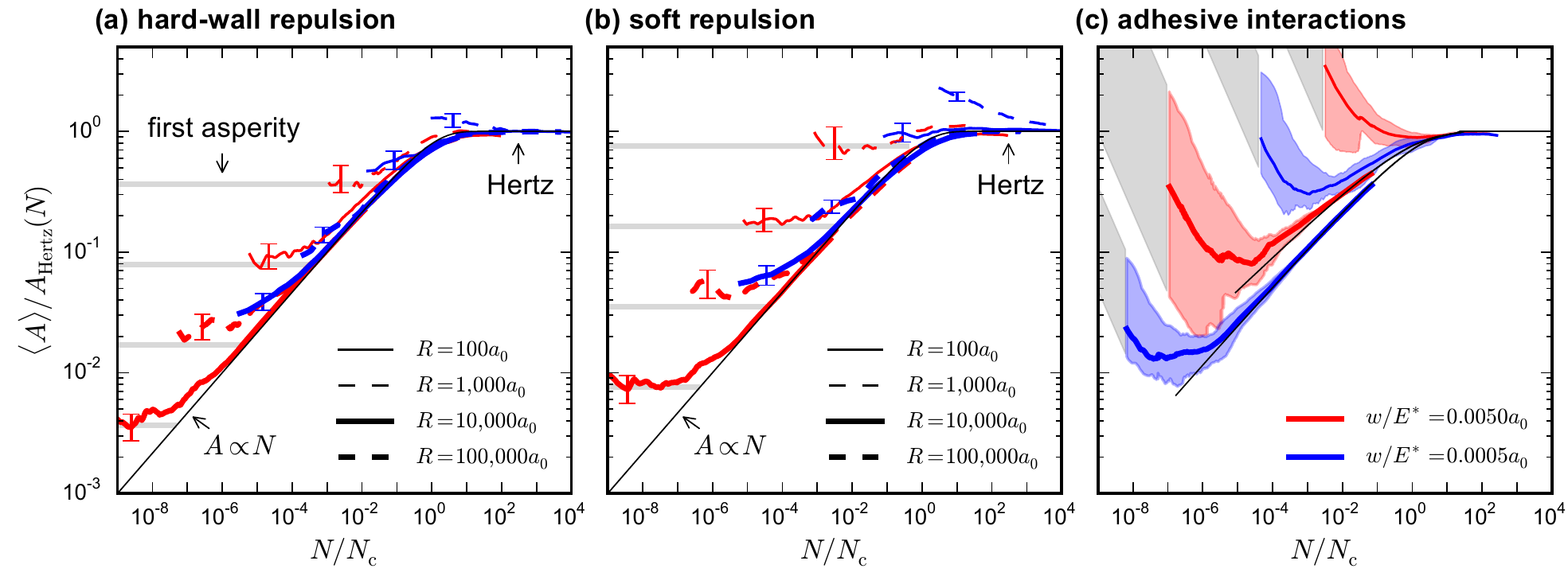}
    \caption{\label{A}
      Dependence of contact area $A$ on normal force $N$.
      The contact area is reported as the fraction of area in contact within the Hertzian area $A_\text{Hertz}$ obtained at the respective normal force.
      The thin black lines are the predictions of Eq.~\eqref{roughsphere}.
      Panel (a) shows results for hard-wall repulsion, while panel (b) shows results for a compliant 9-3 Lennard-Jones potential (Eq. S-7)\cite{supplemental}.
      Data in panels (a) and (b) is shown for $h_{\rm rms}^\prime=0.1$ (red lines) and $h_{\rm rms}^\prime=0.01$ (blue lines).
      Error bars indicate the standard deviation of the five surface realizations probed here.
      Panel (c) shows results for adhesive interactions with the indicated $w/E^*$, $h_{\rm rms}^\prime=0.1$ and sphere radii of $R=1,000a_0$ and $R=100,000 a_0$ (thick and thin lines, respectively).
      The gray and colored shaded regions are bound by the minimum and maximum contact area obtained for our realizations of a rough surface.
      Roughness has Hurst exponent $H=0.8$ and small and large wavelength cutoffs $\lambda_s=4 a_0$ and $\lambda_L=512 a_0$, respectively.
    }
  \end{center}
\end{figure*}

Figure~\ref{A}a shows that this simple scaling assumption (thin solid line) captures numerical results for hard wall interactions over a wide range of sphere radii.
For each load, the area $\langle A\rangle$ is averaged over five realizations of a rough sphere and normalized by $A_\text{Hertz}(N)$.
When the load is normalized by $N_c$, data for different $R$ collapse onto Eq.~\eqref{roughsphere} at intermediate to large loads.
Note that the collapse covers up to $10$ orders of magnitude in both $\langle A\rangle$ and $N$, since $A_\text{Hertz}\propto N^{2/3}$.

Fig.~\ref{A}a reveals yet another scaling regime at the lowest loads where $A/A_\text{Hertz}={\rm const.}\ll 1$.
Here contact is dominated by the first rough peak to make contact.
The peak radius $\rho$ can be estimated from the rms 
curvature $\sqrt{\langle |\nabla^2 h|^2\rangle}$ of the rough surface as~\cite{supplemental}
%
%
\begin{equation}
  \rho=2\langle |\nabla^2 h|^2\rangle^{-1/2}=\lambda_s/(2\pi h^\prime_{\rm rms}) [2(2-H)/(1-H)]^{1/2}
  \label{aspr}
\end{equation}
The gray horizontal lines in Fig.~\ref{A}a show that the result of this calculation gives an excellent agreement with the numerical data at low loads.
Note that Eq.~\eqref{areacross} with $R$ replaced by $\rho$ also describes the crossover from this first contacting asperity to multiasperity contact.
The multiasperity regime where $A \propto N$ is suppressed for small spheres ($\rho \sim R$)
because there is never a statistical number of asperities.

Real interatomic interactions have a finite stiffness and an adhesive tail~\cite{Israelachvili:Book1991}.
A common model for the force between surfaces is
a 9-3 Lennard-Jones interaction that is
truncated at its minimum so it is purely repulsive.
Fig.~\ref{A}b shows results for this model with the
strength of the potential chosen so that the interaction has the
same stiffness as the bulk (see Supporting Section S-II).
We find that the area is generally larger than for the hard-wall case.
At intermediate loads the increase is captured by Eq. (3) with $1/\kappa$
decreased to $1/\kappa_\text{rep}$, consistent
with past studies of nominally flat surfaces with soft repulsion~\cite{Muser:2008p055504,Akarapu:2011p204301,Prodanov:2013p433}.
The finite compliance of the interfacial repulsion leads to variations
in surface separation and supplemental section S-2~\cite{supplemental} provides an analytic expression for the resulting changes in $1/\kappa_\text{rep}$.
Fig.~\ref{A}b shows that the renormalized value of $\kappa_\text{rep}=2.9$ (thin black line labeled $A\propto N$) obtained from this expression agrees well with the numerical data in the linear scaling regime.
The contact of the first asperity is approximately Hertzian.
Compliance increases the effective radius to $\rho^\text{eff}\approx 3\rho$ (gray horizontal lines in Fig.~\ref{A}b).

Finally, we consider longer-range attractive interactions that favor adhesion
between sphere and substrate~\cite{supplemental}.
%
This introduces a new quantity, the work of adhesion $w$ which is the energy gained per unit contact area~\cite{Israelachvili:Book1991,Maugis:Book1999}.
The competition between adhesion and elastic deformation is quantified by the elastocapillary
length $w/E^*$, which is normally much less than an atomic spacing~\cite{Pastewka:2014p3298,roman2010}.
The main effect of adhesion in the intermediate $A\propto N$ scaling regime~\cite{Pastewka:2014p3298} is to approximately renormalize the value of $1/\kappa$ to $1/\kappa_\text{adh}=1/\kappa_\text{rep}-1/\kappa_{\rm att}$.
The value of $1/\kappa_\text{att}$ increases with $w/E^*$ and there is macroscopic adhesion when $\kappa_\text{adh}$ becomes negative.
Using the analytical theory of Ref.~\onlinecite{Pastewka:2014p3298}, we find $1/\kappa_{\rm att}\approx 0.04$ and $1/\kappa_{\rm att}\approx 0.25$ for $w/E^*=0.0005 a_0$ and $w/E^*=0.0050 a_0$, respectively.
These values correspond to van der Waals adhesion ($w\sim 50\,\text{mJ}\,\text{m}^{-2}$) on stiff ceramic substrates ($E^*\sim 300\,\text{GPa}$) and metals ($E^*\sim 30\,\text{GPa}$), respectively.

Renormalizing $\kappa$ works well for the weakly adhesive case where $1/\kappa_{\rm att}\lesssim 1/\kappa_\text{rep}$.
For the strongly adhesive case we replace the expression for the fractional area, Eq.~\eqref{fracarea}, with tabulated values obtained from a reference calculation on nominally flat surfaces and an identical interaction potential.
The thin black lines in Fig.~\ref{A}c show that the result of this procedure is in excellent agreement with our calculations of adhesive spheres, even in strongly adhesive cases.
The influence of the curvature of the macroscopic sphere, and hence the change in nominal contact area, is still captured well by Hertz's expression as long as $\kappa_\text{adh}$ remains positive.

The numerical data for adhesive interactions show that the contact of the first asperity leads to different asymptotic behavior than in the nonadhesive case and much larger variations between different realizations of the random surfaces.
For an analytical description of the asymptotic behavior at low loads, Hertz's expression needs to be replaced by its adhesive counterpart:
The Johnson-Kendall-Roberts (JKR)~\cite{Johnson:1971p301} or Derjaguin-Muller-Toporov (DMT)~\cite{Derjaguin:1975p314} models are commonly used.
These models present bounding limits of continuum contact for soft substrates and large $w$ and stiff substrates and small $w$, respectively.
The general intermediate case can be described by the Maugis-Dugdale solution~\cite{Maugis:1992p243}.
Our numerical data is well-described by simply assuming a constant contact area $A_\text{adh}$ for the first asperity, which all of these models predict at low positive load.
The results of fitting $A_\text{adh}$ individually to our ensemble of five rough surfaces is shown by the gray area in Fig.~\ref{A}c.

For all of the surface interactions in Fig.~\ref{A}, the range of linear scaling ($A \propto N$) decreases as $R/\lambda_s$ and $h'_\text{rms}$ decrease.
Linear scaling requires a statistical number of asperities~\cite{Hyun:2004p026117} and this requires a large separation between $R$ and the size of the first asperity to contact.
The smallest $R=100a_0 \sim 30{\rm nm}$ corresponds to a typical atomic force
microscope tip.
In contrast to some earlier results~\cite{Mo:2009p1116}, we find no significant linear region for tips that are this small, even when the smallest scale of roughness is an atomic diameter and an artificial hard sphere
interaction is used on this scale.
In contrast there is a substantial linear region for $R =10^5a_0 \sim 30\,\mu$m,
which is typical of colloidal probe experiments or characteristic asperity radii inferred from macroscopic friction experiments~\cite{Rabinowicz:Book1995}.

Our results have important consequences for fits to experimental data on a wide range of scales. At low loads and contact areas, recent results for rough flat surfaces can be applied to rough spheres~\cite{perssonreview,Hyun:2004p026117,Campana:2007p38005,Akarapu:2011p204301,Putignano:2012p973,Prodanov:2013p433,Carbone:2008p2555,Pastewka:2014p3298}.
Smooth sphere theories~\cite{Hertz:1881p156,Johnson:1971p301,Derjaguin:1975p314,Johnson:Book1985,Maugis:Book1999,Maugis:1992p243} can be used to fit experimental data in the limit of full contact,
i.e. when the contact radius is larger than the value corresponding to $N_c$, $a/R=(3\pi/4 \kappa) h^{\prime}_\text{rms}$.
Fit values of $w$ will in general be reduced from the intrinsic interaction energy due to the energy cost
of flattening the roughness~\cite{Persson:2001p5597,Mulakaluri2011}.
Excellent fits to JKR theory have been obtained for atomically flat mica and very smooth elastomer surfaces with
$R$ from $1$ to $10\,\text{mm}$ and $a/R$ from 0.001~\cite{Chaudhury:1991p1013} to 0.1~\cite{Horn:1987p480,McGuiggan:2007p5984}.
Many AFM experiments with $R \sim 30$nm are also consistent with DMT or JKR theory~\cite{Carpick:1996,Schwarz:1997p6987,Lessel2013}.
This is consistent with our finding that there is no intermediate $A\propto N$
region for these small tips (Fig. \ref{A}).
Of course the roughness may affect the effective radius that should be used in extracting $w$~\cite{Jacobs:2013p81}.

In summary, we have developed simple analytic expressions for the contact of rough spheres and tested them with large scale simulations with tip radii from $30\,\text{nm}$ to $30\,\mu\text{m}$ over $10$ orders of magnitude in load. Our equations combine established continuum theories for smooth spheres~\cite{Hertz:1881p156,Johnson:1971p301,Derjaguin:1975p314,Johnson:Book1985,Maugis:Book1999,Maugis:1992p243} and recent results for nominally flat but rough surfaces~\cite{perssonreview,Hyun:2004p026117,Campana:2007p38005,Akarapu:2011p204301,Putignano:2012p973,Carbone:2008p2555,Prodanov:2013p433,Pastewka:2014p3298}. The latter are extended from the ideal limit of hard wall interactions to include finite compliance and adhesion which become important at nanometer scales. The results provide detailed predictions for the functional form of the contact area as well as simple equations for estimating which regime describes an experiment or simulation. 
Above a crossover load $N_\text{c}$ (Eq.~\eqref{areacross}), the contact area follows continuum theories for smooth spheres using the macroscopic radius of curvature $R$.
The same theories can be used at very small loads, but with $R$ replaced by the radius $\rho$ of the first asperity to make contact (Eq.~\eqref{aspr}). When $R\gg \rho$ there is an intermediate range of pressures where the area is proportional to load and is consistent with results for rough flat surfaces. A simple criterion describes whether surfaces adhere in this limit~\cite{Pastewka:2014p3298}.

\emph{Acknowledgments –--}
This material is based upon work supported by the
Deutsche Forschungsgemeinschaft under Grant PA 2023/2 and
U. S. National Science Foundation Grants No. DMR-1411144, OCI-0963185 and CMMI-0923018.
We additionally thank the J\"ulich Supercomputing Center for providing computer time (project hfr13) and the Large Scale Data Facility at KIT for providing storage space.


\begin{thebibliography}{48}
\bibitem [{apr()}]{aprime}%
  \BibitemOpen
  \href@noop {} {}\bibinfo {note} {Note that for $a'<\lambda_L$, $h_{\rm rms}$
  depends on the scale $a'$ over which it is measured, $h_{\rm rms}\propto
  a'^H$, where $H$ is the Hurst exponent of the self-affine rough
  surface.}\BibitemShut {Stop}%
\end{thebibliography}
\end{document}